\documentclass[pre,showpacs,preprint]{revtex4}

\usepackage{amsmath}

\usepackage{graphicx}
\usepackage{dcolumn}
\usepackage{bm}
\DeclareMathOperator{\Tr}{Tr}

\begin{document}

\title{Hydrodynamics for a model of a confined quasi-two-dimensional granular gas}
\author{J. Javier Brey, V. Buz\'{o}n, P. Maynar, and M.I. Garc\'{\i}a de Soria}
\affiliation{F\'{\i}sica Te\'{o}rica, Universidad de Sevilla,
Apartado de Correos 1065, E-41080, Sevilla, Spain}
\date{\today }

\begin{abstract}
The hydrodynamic equations for a model of  a confined quasi-two-dimensional gas of smooth inelastic hard spheres are derived from the Boltzmann equation for the model, using a generalization of the Chapman-Enskog method. The heat and momentum fluxes are calculated to Navier-Stokes order, and the associated transport coefficients are explicitly determined as functions of the coefficient of normal restitution and the velocity parameter involved in the definition of the model. Also an Euler transport term contributing to the energy transport equation is considered. This term arises from the gradient expansion of the rate of change of the temperature due to the inelasticity of collisions, and vanishes for elastic systems. The hydrodynamic equations are particularized for the relevant case of a system in the homogeneous steady state. The relationship  with previous works is analyzed.
\end{abstract}

\pacs{45.70.Mg,45.70.-n,05.20.Jj,51.10.+y}

\maketitle

\section{Introduction}
\label{s1}
Granular gases frequently exhibit flows similar to those of normal gases, and for practical purposes these flows are often successfully described by phenomenological hydrodynamic equations \cite{Ha83,Ca90}. The basis for such macroscopic balance equations are in the more fundamental statistical mechanics and kinetic theory descriptions of granular gases. In this context, the idealized model of a granular gas as a monodisperse system of smooth inelastic hard spheres or disks with a constant coefficient of normal restitution has been extensively employed \cite{Du00,Go03}. For this system, hydrodynamic equations to Navier-Stokes order have been derived with expressions for the parameters appearing in them. Starting from the Boltzmann equation for inelastic hard spheres or disks \cite{BDKyS98,ByC01} and also from the revised Enskog theory \cite{GyD99}, the transport coefficients have been evaluated by using an extension of the Chapman-Enskog method. The predictions from the Boltzmann equation have been found to be in good agreement with the values obtained by particle simulation methods in the dilute limit \cite{ByC01}.  Using linear response theory, formal Green-Kubo like expressions for the transport coefficients have been derived for low density granular gases \cite{DyB02,ByR04}, and also for arbitrary densities  \cite{DByL02,BDyB08}. The latter are not tied to any specific kinetic equation,  but their explicit evaluation requires the introduction of some approximations \cite{DByL02}.

In normal fluids, non-equilibrium steady states can be generated by imposing appropriate boundary conditions. Moreover, the control of the boundary conditions permits to keep the gradients of the hydrodynamic fields small, so that the steady state can be studied in the Navier-Stokes domain of the hydrodynamic equations. In granular gases, a new class of steady states shows up. In them, stationarity is reached by an autonomous balance between external constrains and the internal cooling. A typical example is a system under shear flow. There is viscous heating due to the work made on the system at the boundaries. If the system is  a granular gas, stationarity is possible when the viscous heating is compensated by the dissipation due to the inelasticity of collisions. The steady state has uniform density and temperature, and a flow velocity with a linear profile. Due to its macroscopic simplicity, it has been extensively studied \cite{LSJyCh84,JyR88,SGyN96,BRyM97,MGSyB99}. This particular steady state exemplifies two features that are characteristic of hydrodynamic steady states of granular gases. First, to compensate the energy dissipation in collisions, the system must develop spatial gradients generating an energy flux, i.e. it must be inhomogeneous. Second, the energy balance leads to a coupling between gradients and inelasticity, so that the limit of small gradients also implies the quasi-elastic limit. Even more, the above coupling is often non-analytical \cite{SGyN96}, implying that the macroscopic description of the steady state  can never be brought within the range of validity of the Navier-Stokes hydrodynamics in those cases.

An interesting alternative to the kind of steady states of granular gases described above has being attracting increasing interest in the last years \cite{OyU05,RPGRSCyM11,CMyS12,GSVyP11,PGGSyV12}. A granular gas is confined to a quasi-two-dimensional geometry by  placing it between two large parallel plates in the horizontal directions, while the distance between the two plates is smaller than two particle diameters, so the system is actually a monolayer since the particles can not jump on another.  The container is vertically vibrated to inject energy through the collisions of the particles with the top and bottom walls. The two-dimensional dynamics of the system when seen from above is considered. It has been observed that it corresponds to that of a two-dimensional granular fluid. Moreover, the system remains homogeneous under a large range of parameters and it eventually reaches a steady homogeneous state.  Very recently \cite{BRyS13}, an idealized  model has been proposed trying to describe the horizontal dynamics in the above experiment, assuming that the particles are smooth inelastic hard spheres. Then, the projections of the particles on the horizontal plane are described as inelastic hard disks, whose collision rule is modified in order to incorporate a mechanism to transfer the energy  injected vertically  to the horizontal degrees of freedom. In this sense, it can be classified as a collisional model. The new collision rule has a constant coefficient of normal restitution $\alpha$, and contains a characteristic velocity $\Delta$ that is added to each particle in a collision so that the normal component of the relative velocity is increased by $2 \Delta$ in  the collision, independently of the effect of the coefficient of normal restitution.  The methods  of non-equilibrium statistical mechanics and kinetic theory developed for inelastic hard spheres and disks \cite{BDyS97,vNEyB98} have been applied to the model \cite{BGMyB13}.  In this way,  the Boltzmann equation and the revised Enskog theory have been formulated. Moreover, the existence of homogeneous hydrodynamics has been analyzed \cite{BMGyB14}. There is a time regime over which the granular temperature of a homogeneous system obeys a closed (hydrodynamic) equation. In the long time limit, the temperature tends to its steady value. Moreover, it has been shown that the homogeneous relaxation of the temperature of the system presents nonlinear memory effects \cite{BGMyB14}, which can be considered as reminiscent of the Kovacs memory effect occurring in the relaxation towards equilibrium of molecular fluids  \cite{Ko63}. 

The aim of this paper is to derive the hydrodynamic equations to Navier-Stokes order for a dilute confined granular gas as described by the collisional model proposed by Brito {\em et al} \cite{BRyS13}. The starting point will be the Boltzmann kinetic equation and the method to be used a generalization of the Chapman-Enskog procedure. The derivation is based on a special ``normal'' solution of the kinetic equation expanded to low order in the gradients of the hydrodynamic fields. The zeroth order approximation is not a local version, both in space and time, of the distribution function of the homogeneous steady state, but it is based on the distribution describing the homogeneous hydrodynamics. This is an important conceptual and practical difference with the standard application of the Chapman-Enskog  method to molecular systems. Of course, it is also possible to consider states close to a stationary one and carry out, for instance, linear response theory around that state to compute transport properties associated to that particular state.  The ranges of applicability of the results obtained by both methods are clearly different, although there can be a common limit  for the simultaneous validity of both. In particular, the Navier-Stokes shear viscosity of a confined dilute granular gas described by the collisional model in a stationary and uniform Couette flow has been computed by employing linear response theory \cite{SRyB15}. The results obtained here for the shear viscosity will be related with those reported in Ref. \cite{SRyB15}.

The remaining of this paper is organized as follows. In the next section, the Boltzmann equation for the model is given, and the exact balance equations for mass, momentum, and energy are derived from it.  The Chapman-Enskog method for obtaining a ``normal'' solution of the kinetic equation as an expansion in the gradients of the hydrodynamic fields  is described. Results through Navier-Stokes order for the pressure tensor and the heat flux are given.  The associated transport coefficients are shown to obey a complete set of first order differential equations. Some details of the calculations are given in Appendices \ref{ap1} and \ref{ap2}, while the explicit results for the transport coefficients are  presented in Sec.  \ref{s3}. The theory is not restricted to any range of values of the coefficient of normal restitution nor of  the characteristic velocity of the model $\Delta$. The contributions to the transport equations coming from the energy sink term due to the non conservation  of kinetic energy in collisions are also discussed. It is shown that there is a first order in the gradients  contribution, Euler term, which is proportional to the divergence of the velocity field.  The associated transport coefficient  is explicitly evaluated.  
Appendix \ref{ap3} provides an sketch of the calculation of this coefficient. The Euler term does not exist in molecular systems and it is peculiar of inelastic collisions \cite{DByB08,GMyT13,GCyV13}, although it vanishes in the low density limit of granular gases composed of smooth inelastic hard spheres or disks \cite{BDKyS98}.  The expressions of the transport coefficients are particularized for the steady homogeneous state in Sec. \ref{s3}. The peculiarity of this state, in which the temperature is a function of the parameters defining the system \cite{BRyS13,BGMyB13}, leads to much simpler expressions of the coefficients. Further comments and conclusions, as well as the relationship with some previous work for the viscosity transport coefficient, are given in Sec.\ \ref{s4}.

\section{Chapman Enskog solution of the Boltzmann equation}
\label{s2}

The Boltzmann equation obeyed by the one-particle distribution function, $f({\bm r},{\bm v},t)$,  of the model reads \cite{BGMyB13}
\begin{equation}
\label{2.1}
\left( \frac{\partial}{\partial t} + {\bm v} \cdot \frac{\partial }{\partial {\bm r}} \right) f({\bm r},{\bm v},t) =
\int d{\bm v}_{1}\, \overline{T}_{0} ({\bm v},{\bm v}_{1}) f({\bm r},{\bm v},t)  f({\bm r},{\bm v}_{1},t).
\end{equation}
Here, $\overline{T}_{0}$ is the binary collision operator defined by
\begin{equation}
\label{2.2}
\overline{T}_{0} ({\bm v}_{1}, {\bm v}_{2}) \equiv \sigma^{d-1} \int d \widehat{\bm \sigma}\, \left[ \Theta \left( {\bm v}_{12} \cdot \widehat{\bm \sigma} - 2 \Delta \right) \left( {\bm v}_{12} \cdot \widehat{\bm \sigma} - 2 \Delta \right) \alpha^{-2} b_{\bm \sigma}^{-1}(1,2)- \Theta ({\bm v}_{12} \cdot \widehat{\bm \sigma}) ({\bm v}_{12} \cdot \widehat{\bm \sigma})
  \right],
\end{equation}
where $\sigma$ is the diameter of the particles, $d$ the dimension of the system \cite{d}, $\widehat{\bm \sigma}$ is  the unit vector joining the center of the two particles at contact, ${\bm v}_{12} \equiv {\bm v}_{1}-{\bm v}_{2}$ is the relative velocity, $\Theta (x)$ is the Heaviside step function, $\alpha$ is the coefficient of normal restitution defined in the interval $0 < \alpha \leq 1$, $\Delta$ is some positive characteristic speed, and $b_{\bm \sigma}^{-1}(i,j)$ is an operator changing all the velocities ${\bm v}_{i}$ and ${\bm v}_{j}$ to its right into the precollisional values corresponding to a collision between them defined by $\widehat{\bm \sigma}$, i.e.,
\begin{equation}
\label{2.3}
b_{\bm \sigma}^{-1}(i,j){\bm v}_{i} = {\bm v}^{*}_{i} = {\bm v}_{i} - \frac{1+ \alpha}{2 \alpha}\,  {\bm v}_{ij} \cdot \widehat{\bm \sigma}  \widehat{\bm \sigma} + \frac{\Delta \widehat{\bm \sigma}}{\alpha},
\end{equation}
\begin{equation}
\label{2.4}
b_{\bm \sigma}^{-1}(i,j){\bm v}_{j} ={\bm v}^{*}_{j} = {\bm v}_{j} + \frac{1+ \alpha}{2 \alpha}\, {\bm v}_{ij} \cdot \widehat{\bm \sigma}  \widehat{\bm \sigma} - \frac{\Delta \widehat{\bm \sigma}}{\alpha}.
\end{equation}
For arbitrary velocity functions, $a({\bm v}_{i},{\bm v}_{j})$ and $b({\bm v}_{i},{\bm v}_{j})$, it is \cite{BGMyB13}
\begin{equation}
\label{2.5}
\int d{\bm v}_{i} \int d {\bm v}_{j}\, b({\bm v}_{i},{\bm v}_{j}) \overline{T}_{0}({\bm v}_{i}, {\bm v}_{j}) a({\bm v}_{i}, {\bm v}_{j})=\int d{\bm v}_{i} \int d {\bm v}_{j}\, a({\bm v}_{i},{\bm v}_{j}) T_{0}({\bm v}_{i}, {\bm v}_{j}) b({\bm v}_{i}, {\bm v}_{j}),
\end{equation}
where
\begin{equation}
\label{2.6}
T_{0}({\bm v}_{i},{\bm v}_{j}) \equiv \sigma^{d-1} \int d \widehat{\bm \sigma}\, \Theta (-{\bm v}_{ij} \cdot \widehat{\bm \sigma} ) |{\bm v}_{ij} \cdot \widehat{\bm \sigma}| \left[ b_{\bm \sigma} (i,j)-1 \right].
\end{equation}
The operator $b_{\bm \sigma} (i,j)$ is the inverse of  $b_{\bm \sigma}^{-1} (i,j)$, i.e. it changes ${\bm v}_{i}$ and ${\bm v}_{j}$ into their postcollisional values, ${\bm v}^{\prime}_{i}$ and ${\bm v}^{\prime}_{j}$, given by
\begin{equation}
\label{2.7}
 b_{\bm \sigma}(i,j){\bm v}_{i}= {\bm v}^{\prime}_{i} = {\bm v}_{i}- \frac{1+ \alpha}{2}  {\bm v}_{ij} \cdot \widehat{\bm \sigma}  \widehat{\bm \sigma} + \Delta \widehat{\bm \sigma},
\end{equation}
\begin{equation}
\label{2.8}
 b_{\bm \sigma}(i,j){\bm v}_{j} = {\bm v}^{\prime}_{j} = {\bm v}_{j}+ \frac{1+ \alpha}{2}  {\bm v}_{ij} \cdot \widehat{\bm \sigma} \widehat{\bm \sigma} - \Delta \widehat{\bm \sigma}.
\end{equation}
The kinetic energy change in a collision is
\begin{equation}
\label{2.9}
e^{\prime}_{ij}-e_{ij} = m\left[ \Delta^{2} - \alpha \Delta  {\bm v}_{ij} \cdot \widehat{\bm \sigma} - \frac{1-\alpha^{2}}{4}\,  ({\bm v}_{ij} \cdot \widehat{\bm \sigma} )^{2} \right],
\end{equation}
with $m$ being the mass of a particle. Using the identity (\ref{2.5}) it is easily found that
\begin{equation}
\label{2.10}
\int d{\bm v} \int d {\bm v}_{1}\, \overline{T}_{0}({\bm v},{\bm v}_{1} ) f({\bm r},{\bm v},t)f({\bm r},{\bm v}_{1},t)=0,
\end{equation}
\begin{equation}
\label{2.11}
\int d{\bm v} \int d {\bm v}_{1}\,  {\bm v} \overline{T}_{0}({\bm v},{\bm v}_{1} ) f({\bm r},{\bm v},t)f({\bm r},{\bm v}_{1},t)=0,
\end{equation}
reflecting the conservation of the number of particles and the momentum, respectively. On the other hand, it is
\begin{equation}
\label{2.12}
\int d{\bm v} \int d {\bm v}_{1}\,  \frac{m  v^{2}}{2} \overline{T}_{0}({\bm v},{\bm v}_{1} ) f({\bm r},{\bm v},t)f({\bm r},{\bm v}_{1},t)= \omega[f,f].
\end{equation}
The term $\omega[f,f]$ provides the rate of energy change due to the inelasticity of collisions, and the functional $\omega[f,h]$ is
\begin{eqnarray}
\label{2.13}
\omega[f,h] & \equiv & \frac{\pi^{(d-1)/2} m \sigma^{d-1}}{2} \int d {\bm v}_{1} \int d {\bm v}_{2}\, f({\bm r},{\bm v}_{1},t) h ({\bm r}, {\bm v}_{2},t) \nonumber \\
&& \times \left[ \frac{ \Delta^{2} v_{12}}{\Gamma \left( \frac{d+1}{2}  \right)}+ \frac{\pi^{1/2} \alpha \Delta v_{12}^{2}}{2 \Gamma \left( \frac{d+2}{2} \right)}- \frac{(1- \alpha^{2})v_{12}^{3}}{4 \Gamma \left( \frac{d+3}{2} \right)} \right].
\end{eqnarray}
The macroscopic number of particles density, $n({\bm r},t)$, flow velocity, ${\bm u}({\bm r},t)$, and granular temperature, $T({\bm r},t)$, are defined from the one-particle distribution function in the usual way,
\begin{equation}
\label{2.14}
n({\bm r},t) \equiv \int d{\bm v}\, f({\bm r},{\bm v},t),
\end{equation}
\begin{equation}
\label{2.15}
n({\bm r},t) {\bm u}({\bm r},t) \equiv \int d{\bm v}\,{\bm v}  f({\bm r},{\bm v},t),
\end{equation}
\begin{equation}
\label{2.16}
\frac{d}{2} n({\bm r},t) T({\bm r},t) \equiv \int d{\bm v}\,\frac{m V^{2}}{2}\,  f({\bm r},{\bm v},t),
\end{equation}
where ${\bm V} ({\bm r},t)= {\bm v}- {\bm u} ({\bm r},t) $ is the velocity of the particle relative to the flow field. Balance equations for the above fields follow by taking velocity moments in the Boltzmann equation, Eq. (\ref{2.1}),
\begin{equation}
\label{2.17}
\frac{\partial n}{\partial t} +  {\bm \nabla} \cdot \left( n {\bm u} \right) =0,
\end{equation}
\begin{equation}
\label{2.18}
\frac{\partial {\bm u}}{\partial t} +{\bm u} \cdot {\bm \nabla} {\bm u} + (mn)^{-1} {\bm \nabla}
\cdot \sf{P}=0,
\end{equation}
\begin{equation}
\label{2.19}
\frac{\partial T}{\partial t} +{\bm u} \cdot {\bm \nabla}T + \frac{2}{nd} \left( {\sf P} : {\bm \nabla} {\bm u} + {\bm \nabla} \cdot {\bm J}_{q} \right) = - \zeta T.
\end{equation}
In the above equations, the pressure tensor, ${\sf P}$, and the heat flux, ${\bm J}_{q}$, are defined by
\begin{equation}
\label{2.20}
{\sf P}({\bm r},t) \equiv m \int d{\bm v}\, {\bm V}({\bm r},t) {\bm V}({\bm r},t) f ({\bm r},{\bm v},t)
\end{equation}
and
\begin{equation}
\label{2.21}
{\bm J}_{q}({\bm r},t) \equiv \frac{m}{2} \int d{\bm v}\,  V^{2}({\bm r},t) {\bm V}({\bm r},t) f({\bm r},{\bm v},t),
\end{equation}
respectively. In addition, Eq. (\ref{2.19}) contains the rate of change of the temperature, $\zeta({\bm r},t)$, due to the inelasticity of collisions, whose expression is
\begin{equation}
\label{2.22}
\zeta ({\bm r},t) \equiv  -\frac{2}{n({\bm r},t) T({\bm r},t) d}\, \omega[f,f].
\end{equation}
The minus sign has been introduced by analogy with a system of smooth inelastic hard spheres or disks, but in the present context it does not presuppose that $\zeta$ is (semi)defined positive \cite{Ha83}. 

To close the balance equations (\ref{2.17})-(\ref{2.19}), it is necessary to express the fluxes and the temperature change rate in terms of the macroscopic fields, by means of some constitutive relations. To accomplish this, the Chapman-Enskog theory \cite{McL89} assumes the existence of a normal solution of the Boltzmann equation, i.e. a solution in which all the dependence of the distribution function on position and time occurs through its functional dependence on the macroscopic fields $n$, ${\bm u}$, and $T$,
\begin{equation}
\label{2.23}
f({\bm r},{\bm v},t) = f[{\bm v}|n, {\bm u}, T].
\end{equation}
Next, it is assumed that the space and time variations of the fields are small, so that the functional dependence of the distribution function on the fields can be localized in space and time by means of an expansion in gradients.  Then, the distribution function is expressed as a power series expansion in a formal uniformity parameter $\epsilon$,
\begin{equation}
\label{2.24}
f=f^{(0)}+ \epsilon f^{(1)}+ \epsilon^{2} f^{(2)}+ \cdots.
\end{equation}
Since the aim is to generate a gradient expansion, a factor of $\epsilon$ is assigned to every gradient operator. Moreover, the Chapman-Enskog method uses the multiple-scale perturbation theory  \cite{Sch91}. In practice, this is done by using  the expansion in Eq. (\ref{2.24}) into the definition of the fluxes and the dissipation rate $\zeta$. Then the resulting expansions are introduced into the macroscopic balance equations to get an identification of the time derivatives of the macroscopic fields in the form of an expansion in the uniformity parameter,
\begin{equation}
\label{2.25}
\frac{\partial}{\partial t}= \partial_{t}^{(0)} + \epsilon \partial_{t}^{(1)} + \epsilon^{2} \partial_{t}^{(2)} + \cdots.
\end{equation}
Details of the application of the method are given in Appendices \ref{ap1}, \ref{ap2}, and \ref{ap3}. To first order in the gradients, the pressure tensor and heat flux are given by
\begin{equation}
\label{2.26}
{\sf P} = n T {\sf I} - \eta \left[ {\bm \nabla} {\bm u} + ( {\bm \nabla} {\bm u})^{+} - \frac{2}{d} {\bm \nabla} \cdot {\bm u} {\sf I} \right],
\end{equation} 
\begin{equation}
\label{2.27}
{\bm J}_{q}= - \kappa {\bm \nabla}T- \mu {\bm \nabla }n,
\end{equation}
where ${\sf I}$ is the unit tensor in $d$ dimensions, $({\bm \nabla} {\bm u})^{+}$ is the transposed of ${\bm \nabla} {\bm u}$,    $\eta$ is the coefficient of shear viscosity, $\kappa$ the heat conductivity, and $\mu$ a new coefficient coupling the heat flux and the density gradient, which is peculiar of inelastic collisions \cite{BMyD96}. To distinguish  between the two energy transport coefficients, sometimes $\kappa$ is referred to as the thermal heat conductivity and $\mu$ as the diffusive heat conductivity. The transport coefficients are determined  by the  normal solutions of the first order differential equations
\begin{equation}
\label{2.28}
\frac{\overline{\zeta}^{(0)} \Delta^{*}}{2} \frac{\partial \overline{\eta}}{\partial \Delta^{*}} + \left(\overline{\nu}_{\eta} - \frac{\overline{\zeta}^{(0)}}{2} \right) \overline{\eta}= \frac{2^{5/2} \pi^{\frac{d-1}{2}}}{(d+2) \Gamma \left( d/2 \right)}\, ,
\end{equation}
\begin{equation}
\label{2.29}
\frac{\overline{\zeta}^{(0)} \Delta^{*}}{2} \frac{\partial \overline{\kappa}}{\partial \Delta^{*}} + \left( \overline{\nu}_{\kappa}+ \frac{\Delta^{*}}{2}  \frac{\partial \overline{\zeta}^{(0)}}{\partial \Delta^{*}} - 2 \overline{\zeta}^{(0)}\right) \overline{\kappa} = \frac{2^{5/2} (d-1) \pi^{\frac{d-1}{2}} }{d(d+2) \Gamma \left( d/2 \right)}\, \left( 1+2a_{2}- \frac{\Delta^{*}}{2} \frac{\partial a_{2}}{\partial \Delta^{*}}\ \right),
\end{equation}
\begin{equation}
\label{2.30}
\frac{\overline{\zeta}^{(0)} \Delta^{*}}{2} \frac{\partial \overline \mu}{\partial \Delta^{*}} + \left( \overline{\nu}_{\mu} - \frac{3 \overline\zeta^{(0)}}{2} \right) \overline{\mu}
- \overline{\zeta}^{(0)} \overline{\kappa} = \frac{2^{5/2} (d-1) \pi^{\frac{d-1}{2}} a_{2}}{d(d+2) \Gamma \left( d/2 \right)}\, .
\end{equation}
In these equations, $\Delta^{*} \equiv  \Delta \left( m/2T \right)^{1/2}$ and dimensionless transport coefficients have been introduced. They are defined by
\begin{equation}
\label{2.31}
\overline{\eta} \equiv \frac{\eta}{\eta_{0}}, \quad \overline{\kappa} \equiv \frac{\kappa}{\kappa_{0}}
,  \quad \overline{\mu} \equiv \frac{n \mu }{T \kappa_{0}},
\end{equation}
 where
 \begin{equation}
 \label{2.32}
 \eta_{0}= \frac{2+d}{8} \Gamma \left( d/2 \right) \pi^{-\frac{d-1}{2}} \left(mT \right)^{1/2} \sigma^{-(d-1)},
 \end{equation}
 and
 \begin{equation}
 \label{2.33}
 \kappa_{0}= \frac{d(d+2)^{2}}{16(d-1)}\, \Gamma \left( d/2 \right) \pi^{-\frac{d-1}{2}} \left( \frac{T}{m} \right)^{1/2} \sigma^{-(d-1)}
 \end{equation}
 are the shear viscosity and the (thermal) heat conductivity, respectively,  of a molecular gas described by the Boltzmann equation, with the Boltzmann constant set equal to unity. The dimensionless functions introduced in Eqs. (\ref{2.28})-(\ref{2.30}) are
 \begin{equation}
 \label{2.34}
 \overline{\zeta}^{(0)} \equiv \frac{\zeta^{(0)}}{n \sigma^{d-1}} \, \left( \frac{m}{2T} \right)^{1/2}, \quad
 \overline{\nu}_{\eta} \equiv \frac{\nu_{\eta}}{n \sigma^{d-1}} \, \left( \frac{m}{2T}\right)^{1/2}, \quad \overline{\nu}_{\kappa} = \overline{\nu}_{\mu}
  \equiv \frac{\nu_{\kappa}}{n \sigma^{d-1}} \, \left( \frac{m}{2T}\right)^{1/2}.
\end{equation}
The expression of the zeroth order rate of change of the temperature, $\zeta^{(0)}$,  is given in Eq.\ (\ref{ap1.13}), while the frequencies $\nu_{\eta}$ and $\nu_{\kappa}$ are given in Eqs. ({\ref{ap2.5}) and (\ref{ap2.6}), respectively. Some details of the calculations are shown in Appendices \ref{ap1} and {\ref{ap2}. Finally, Eqs. (\ref{2.28})-(\ref{2.30}) have been obtained by considering the distribution function in the called first Sonine approximation, in which the deviation of the one-particle distribution function of the gas from the Gaussian are characterized by the coefficient $a_{2}$, given by the normal solution of the ordinary differential equation (\ref{ap1.14}). Moreover, the same kind of approximation has been considering upon evaluating the hydrodynamic fluxes. This is the usual approximation to obtain explicit expressions for the transport coefficients of a gas with elastic collisions and  there is no reason to question its accuracy here as well. Actually, it has been shown to lead to quantitatively right approximations in the case of a system of smooth inelastic hard spheres or disks \cite{ByC01}.

\section{Euler and Navier-Stokes transport coefficients}
\label{s3}
As a consequence of the confinement of the fluid and its description by means of a modified hard collision,  there is a  contribution to the hydrodynamic equations of the rate of change of the temperature $\zeta ({\bm r},t)$ of first order in the gradients, namely it is (see Eq.\ (\ref{ap1.32}))
\begin{equation}
\label{3.1}
\zeta^{(1)} ({\bm r},t) = \zeta_{1} {\bm \nabla} \cdot {\bm u}.
\end{equation}
The Euler transport coefficient $\zeta_{1}$  represents dissipation due to the inelastic character of collisions proportional to ${\bm \nabla} \cdot  {\bm u}$. It has no analogue for elastic fluids, where the Euler hydrodynamics (first order in the gradients of the fields)  is referred to as the  ``perfect fluid''  equations, since there is not dissipation in that limit. The expression derived here vanishes in the limit $\Delta^{*} \rightarrow 0$, as a consequence of symmetry considerations \cite{BDKyS98}, i.e. there is no dissipation to Euler order in a dilute gas of smooth inelastic hard spheres or disks. On the other hand, a term of the form given in Eq. (\ref{3.1}) is present in the hydrodynamic equations even in systems of smooth particles if density effects are considered \cite{GyD99,DByB08,BDyB08}. The calculation of $\zeta_{1}$ using the Chapman-Enskog procedure, requires to determine $f^{(1)}$. The details of the calculation are given in Appendix \ref{ap3}, where the first Sonine polynomial expansion is again employed. The same approximation was used above  to compute the Navier-Stokes transport coefficients.  In Fig.\ \ref{fig1},  the dimensionless coefficient $ \zeta_{1}$  
is plotted as a function of the speed parameter for two values of the coefficient of normal restitution, namely $\alpha=0.8$ and $\alpha=0.9$. The value of the transport coefficient in the  homogenous steady state corresponding to each value of $\alpha$ is indicate in the figure. To find the values of the transport coefficient a coupled of first order differential equations have to be solved, namely Eqs. (\ref{ap1.14}) and (\ref{ap3.3}). The curves reported in the figure correspond to the hydrodynamic solutions of the equations, i.e. they are identified independently of the initial conditions. The method is described below in detail for the shear viscosity coefficient and it will be not  be  discussed now.

The coefficient $\zeta_{1} $ is the contribution of the energy source in collisions to what would physically constitute the effects of the hydrostatic pressure at the Euler order. If a small element of the confined granular gas is considered, then the pressure that the fluid element exerts on its boundaries is decreased or increased by the energy lost locally in collisions. At the level of linear hydrodynamics, the pressure and the Euler dissipative term are indistinguishable in their physical implications \cite{BDyB08}. 

\begin{figure}
\includegraphics[scale=0.6,angle=0]{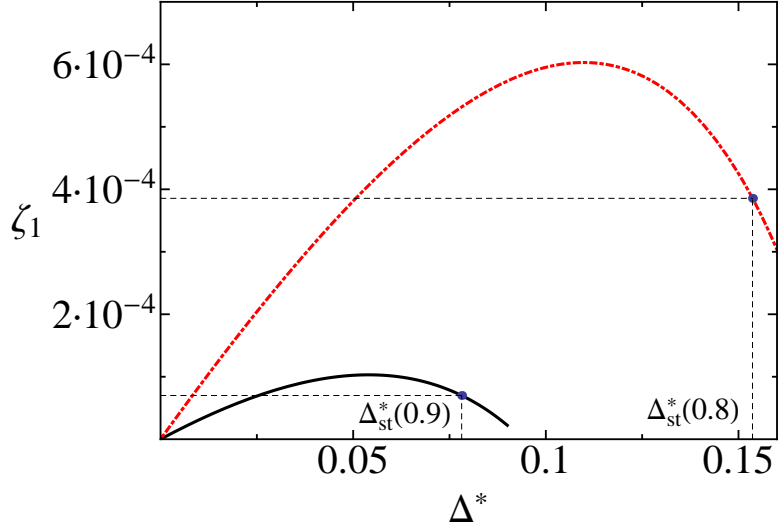}
\caption{(Color online) Dimensionless Euler transport coefficient  $\zeta_{1}$ as a function of dimensionless characteristic speed $\Delta^{*}$ in a two-dimensional system. The (red) dashed line is for a coefficient of normal restitution $\alpha=0.8$ and the (black) solid line is for $\alpha=0.9$. The (blue) dots indicate the values of the transport coefficient in each of the two steady states.}
\label{fig1}
\end{figure}

To compute the Navier-Stokes transport coefficients, Eqs.\ (\ref{2.28})-(\ref{2.30}) have to be solved. In the equations, the hydrodynamic expression of the second Sonine coefficient, $a_{2}$, has to be used. The latter is obtained by numerically solving Eq. (\ref{ap1.14}) as a function of $\Delta^{*}$ for a fixed value of $\alpha$, and a given initial condition $a_{2}(\alpha, \Delta^{*}_{0})=a_{2,0}$. It is seen that all the trajectories converge quite fast towards a universal curve, identified as the hydrodynamic expression of the second Sonine coefficient \cite{BMGyB14}. A similar method has been employed here to generate the hydrodynamic transport coefficients. Actually, what has been done is to simultaneously solve the equation for $a_{2}$ and those for the transport coefficients. Since the rate of variation of the temperature vanishes in the steady state, the equations for the transport coefficients have a singularity at the steady value  $\Delta^{*}= \Delta^{*}_{st}$. Therefore, in the numerical simulations, trajectories have been generated starting from both $\Delta^{*} _{0}> \Delta^{*}_{st}$ and with $\Delta^{*}_{0} < \Delta^{*}_{st}$. The hydrodynamic solution giving the expression of the transport coefficient is the common part of all the numerical solutions. As an example, the numerical results obtained for the dimensionless coefficient $\overline{\eta}$ in a two dimensional system with $\alpha=0.9$ are shown in Fig.\ \ref{fig2}. All the numerical trajectories converge towards the same curve, then forgetting the initial conditions used to generate them. This is consistent with the existence of a hydrodynamic shear viscosity being a function of only the local hydrodynamic fields, but not of the previous history or some initial values. In the particular case shown in Fig.\  \ref{fig2}, several initial values of the viscosity parameter corresponding to $\Delta^{*}= 0.005$ and $\Delta^{*}=10$ have been employed. The curves tend to converge quite fast  in the range $0 \leq \Delta^{*} \lesssim 0.2$. For $\Delta^{*} \gtrsim 0.2$, the dependence of the solution of the differential equation on the initial value of $\overline{\eta}$ used for $\Delta^{*}=\Delta^{*}_{0}$ is rather strong and much more intensive numerical simulations would be needed to identify the value of the hydrodynamic shear viscosity. The two particular solutions  drawn in Fig.\ \ref{fig2} for $\Delta^{*}>\Delta^{*}_{st}$ correspond to $\overline{\eta}(\Delta^{*}=10)=100 $ and  $\overline{\eta}(\Delta^{*}=10)=0$, respectively, while the third plotted particular solution has been obtained with the initial condition  $\overline{\eta}(\Delta^{*}=0.005)= 10$. Results obtained with other initial conditions can not be distinguished from the normal curve on the scale of the figure.

\begin{figure}
\includegraphics[scale=0.6,angle=0]{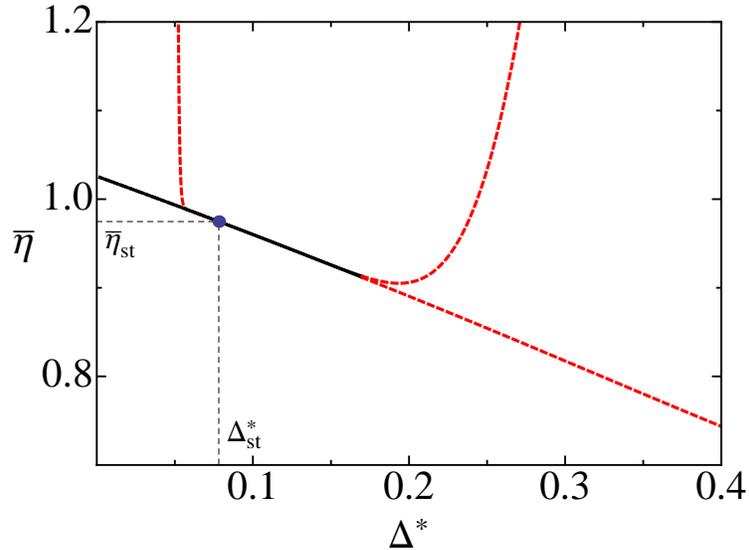}
\caption{(Color online) Adimensionalized  quantity $\overline{\eta}$ as a function of the dimensionless characteristic speed $\Delta^{*}$ in a two-dimensional system with $\alpha=0.9$. The (red) dashed lines correspond to  numerical solutions of Eq.\ (\ref{2.28}) obtained by using different initial conditions, i.e. different values for the pair $\Delta^{*}_{0}, \overline{\eta}(\Delta^{*}_{0})$. The (black) solid line is the universal curve to which all the solutions converge. This is precisely the function identified as the dimensionless hydrodynamic shear viscosity. Also indicated in the figure is the steady value of $\Delta^{*}$, denoted by $\Delta^{*}_{st}$ and the shear viscosity of the steady state, $\overline{\eta}_{st}$ .}
\label{fig2}
\end{figure}

In Figs.\  \ref{fig3} - \ref{fig5} the coefficients of shear viscosity, $\overline{\eta}$, (thermal) heat conductivity, $\overline{\kappa}$, and diffusive heat conductivity, $\overline{\mu}$, are plotted as a function of the dimensionless characteristic speed for a two-dimensional system. Two values of the  coefficient of normal restitution have been considered, namely $\alpha=0.8$ and  $\alpha=0.9$. The reported curves correspond to the hydrodynamic transport coefficients and have been obtained by the same method as described above for the shear viscosity. The values of the several transport coefficients in the steady state are indicated.

\begin{figure}
\includegraphics[scale=0.6,angle=0]{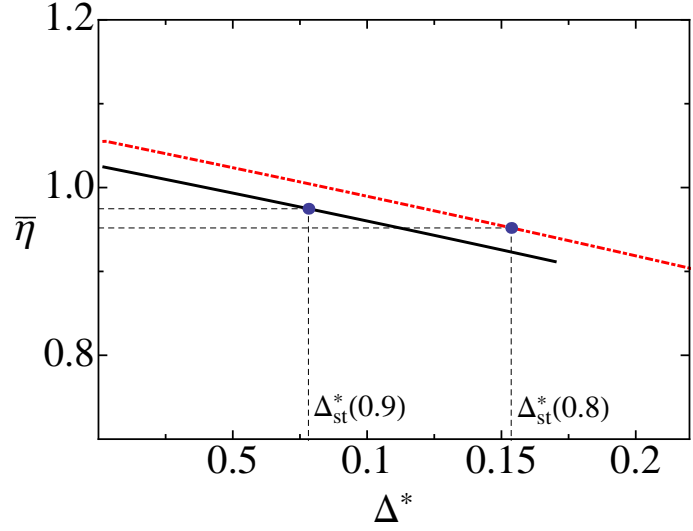}
\caption{(Color online) Adimensionalized shear viscosity $\overline{\eta}$ of a two-dimensional system as a function of the dimensionless  speed parameter $\Delta^{*}$. The (red) dashed curve is for $\alpha=0.8$ and the (black) solid line is for $\alpha=0.9$. The (blue) dots indicate the steady state values in each case.}
\label{fig3}
\end{figure}

\begin{figure}
\includegraphics[scale=0.6,angle=0]{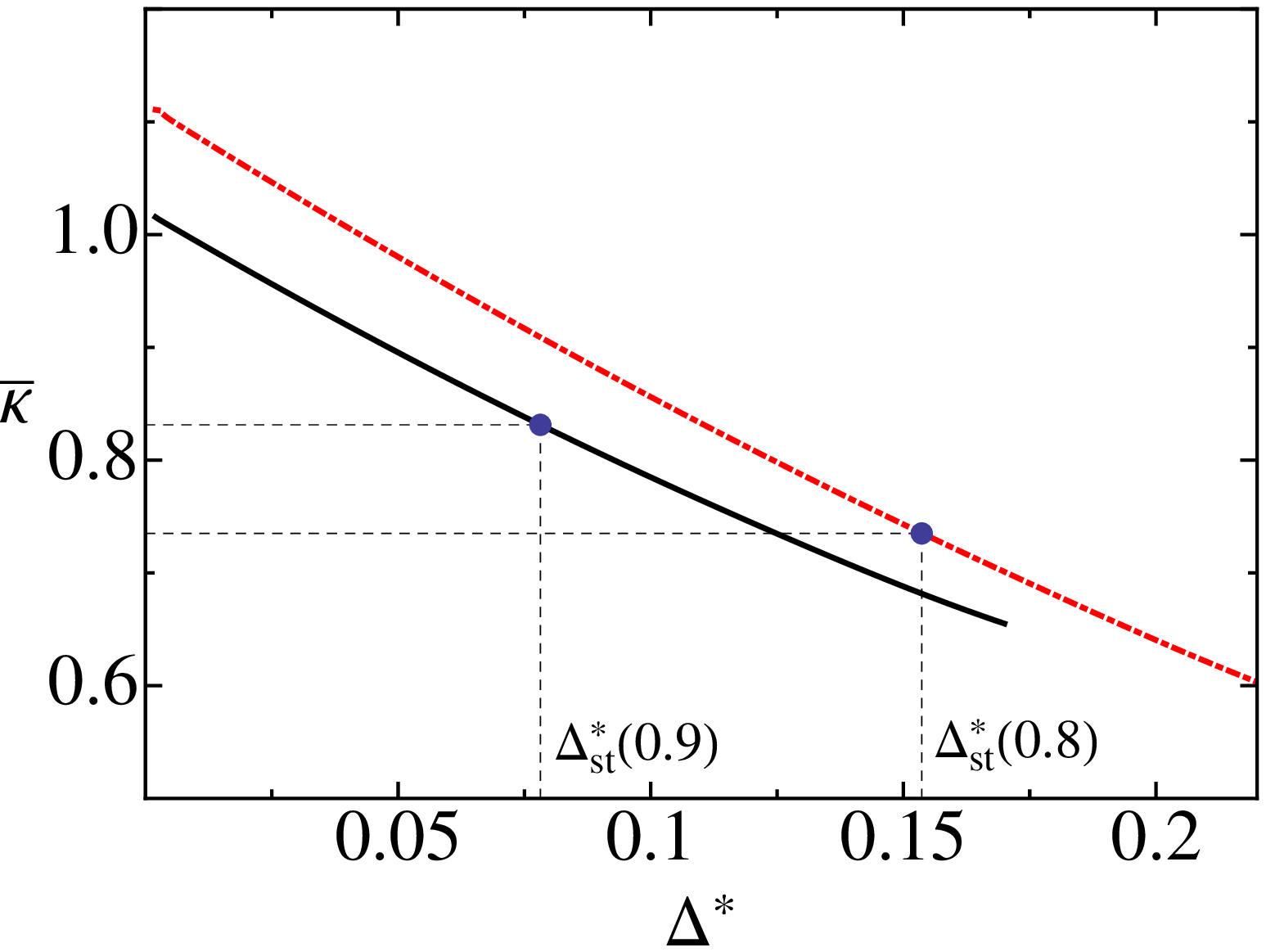}
\caption{(Color online) Adimensionalized (thermal) heat conductivity $\overline{\kappa}$ of a two-dimensional system as a function of the dimensionless  speed parameter $\Delta^{*}$. The (red) dashed curve is for $\alpha=0.8$ and the (black) solid line is for $\alpha=0.9$. The (blue) dots indicate the steady state values in each case. }
\label{fig4}
\end{figure}

\begin{figure}
\includegraphics[scale=0.6,angle=0]{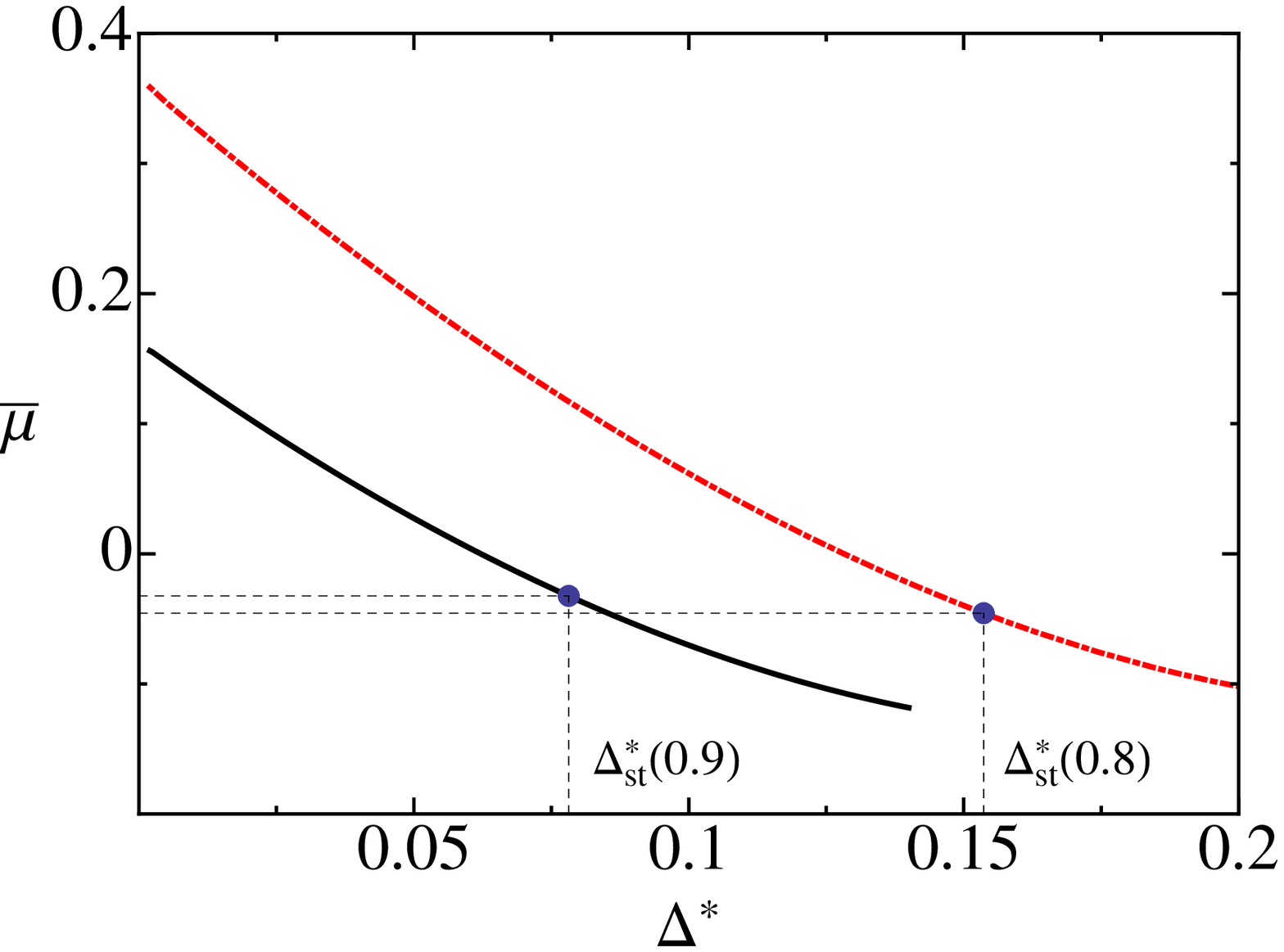}
\caption{(Color online) Adimensionalized diffusive heat conductivity $\overline{\mu}$ as a function of the dimensionless  speed parameter $\Delta^{*}$ for a two dimensional system. The (red) dashed curve is for $\alpha=0.8$ and the (black) solid line is for $\alpha=0.9$. The (blue) dots indicate the steady state values in each case.}
\label{fig5}
\end{figure}

It is observed that the three Navier-Stokes transport coefficients are monotonically decreasing functions of the speed parameter $\Delta^{*}$ for the two values of the restitution coefficient $\alpha$ considered in the figures. A similar behavior was found for other values of $\alpha$. The coefficient of diffusive heat conductivity $\mu$, becomes even negative for large enough values  of $\Delta^{*}$. Notice that  this does not seem to imply the violation of any fundamental physical law or be incompatible with any  physical symmetry. Nevertheless, it is quite possible that the exact value of $\Delta^{*}$ at which the change in sign of $\mu$ occurs be a consequence of the approximations made and, in particular, of the first Sonine approximation. When the prediction for $\mu$ in this approximation is rather small, it is evident that higher order corrections might become relevant.

\section{Transport coefficients in the homogeneous steady state}
\label{s4}
A particularly relevant state of the confined granular gas is the homogeneous steady state. Stationarity of the temperature implies that the rate of change of the temperature vanishes, i.e. it is
\begin{equation}
\label{4.1}
\overline{\zeta}^{(0)}(\Delta^{*}_{st})=0.
\end{equation}
Then, particularization of Eq.\ (\ref{ap3.3}) for the steady state yields
\begin{eqnarray}
\label{4.1a}
\zeta_{1,st} &= & 2 \Delta^{*}_{st} \left( \frac{\partial a_{2}}{\partial \Delta^{*}} \right)_{\Delta^{*}= \Delta^{*}_{st}}\  \overline{\zeta}_{1}(\Delta^{*}_{st})  \nonumber \\&& \times \left\{ \frac{8 \chi(\Delta^{*}_{st})}{d(d+2)} + d \overline{\zeta}_{1}(\Delta^{*}_{st})  \left[ 4(a_{2,st}+1)- \Delta^{*}_{st} \left( \frac{\partial a_{2}}{\partial \Delta^{*}} \right)_{\Delta^{*}  = \Delta^{*}_{st}}\ \right] \right\}^{-1}.
\end{eqnarray}
Similarly, particularization of Eqs. (\ref{2.28})-(\ref{2.30}) for the steady state lead to explicit expressions for the Navier-Stokes transport coefficients for this state,
\begin{equation}
\label{4.2}
\overline{\eta}_{st}= \frac{2^{5/2} \pi^{\frac{d-1}{2}}}{(d+2) \Gamma \left( d/2 \right)}\, \overline{\nu}_{\eta}^{-1}(\Delta^{*}_{st},a_{2,st}),
\end{equation}
\begin{eqnarray}
\label{4.3}
\overline{\kappa}_{st} & = & \frac{2^{5/2} (d-1) \pi^{\frac{d-1}{2}} }{d(d+2) \Gamma \left( d/2 \right)}\, \left[ 1+2a_{2,st}- \frac{\Delta^{*}_{st}}{2} 
\left( \frac{\partial a_{2}}{\partial \Delta^{*}} \right)_{\Delta^{*}= \Delta^{*}_{st}}\ \right]  \nonumber \\
&& \times \left[ \overline{\nu}_{\kappa} (\Delta^{*}_{st},a_{2,st})+ \frac{\Delta^{*}_{st}}{2}  \left( \frac{\partial \overline{\zeta}^{(0)}}{\partial \Delta^{*}} \right)_{\Delta^{*}= \Delta^{*}_{st}} \right]^{-1}, 
\end{eqnarray}
\begin{equation}
\label{4.4}
\overline{\mu}_{st}= \frac{2^{5/2} (d-1) \pi^{\frac{d-1}{2}} a_{2,st}}{d(d+2) \Gamma \left( d/2 \right) \overline{\nu}_{\mu}( \Delta^{*}_{st},a_{2,st})}\, .
\end{equation}
The frequencies $\overline{\nu}_{\eta}$ and $\overline{\nu}_{\kappa}= \overline{\nu}_{\mu}$ are defined in Eqs.\ (\ref{2.34}). Moreover, the calculation of $a_{2}(\Delta^{*})$ for $\Delta^{*}$ in the vicinity of $\Delta^{*}_{st}$ can be carried out in an efficient and quite accurate way by noting that near $\Delta^{*}_{st}$ it is $|\partial a_{2} / \partial \Delta^{*} | \ll 1$ (see, for instance, Fig. 9 in Ref. \cite{BGMyB14}). Then, near the steady state, Eq. (\ref{ap1.14}) yields
\begin{equation}
\label{4.5}
a_{2} \approx - \frac{B_{0}+4A_{0}}{B_{1}+4(A_{0}+A_{1})}.
\end{equation}
The expressions of $A_{0}$, $A_{1}$, $B_{0}$, and $B_{1}$ are given in Eqs. (\ref{ap1.15})-(\ref{ap1.18}).

It is now a simple task to evaluate the Euler and Navier-Stokes transport coefficients in the steady state. They are plotted in  Figs. \ref{fig6}-\ref{fig9} as a function of the coefficient of normal restitution $\alpha$. The four coefficients present a clear dependence with the inelasticity. The Euler transport coefficient is a monotonic decreasing function of the coefficient of normal restitution, while the dimensionless shear viscosity monotonically increases with the value of $\alpha$. This latter behavior is consistent with molecular dynamics simulation results reported both for dilute \cite{SRyB15} and moderately dense  systems
\cite{BRyS13}. Moreover, the dependence of the viscosity on the coefficient of normal restitution is clearly nonlinear, again in agreement with the simulations for dilute systems. It is worth to remind that in (non-confined) dilute granular gases of smooth inelastic hard spheres \cite{BDKyS98,ByC01} the viscosity decreases as the coefficient of normal restitution increases, and that in a stochastic thermostat model it has been found to be a non-monotonic function of the inelasticity \cite{GMyT13,GCyV13}.  On the other hand, the dependence on the restitution coefficients of the two transport coefficients associated with the heat flux is not monotonic in the homogeneous steady state of the model discussed here, exhibiting both a minimum. Moreover, the coefficient of diffusive heat conductivity $\mu_{st}$ is negative in the whole range of values of $\alpha$, while it is always positive in a dilute non-confined gas of inelastic hard spheres or disks. Notice that the dependence of the steady transport coefficients on the inelasticity of the system is quite strong. In particular, the (thermal) heat conductivity  for $\alpha=0.8$ is about $25\%$ smaller than its elastic limit value.

In any case, when interpreting the results in Figs. \ref{fig6}-\ref{fig9}, it must be kept in mind that the bare transport coefficients have been scaled with a function of the temperature of the steady state, and that this temperature is in turn a function of the coefficient of normal restitution (and the velocity parameter of the model $\Delta$). As a consequence, it is not possible to deduce the general expressions of the transport coefficients of the model to Navier-Stokes order from their form in the steady homogeneous  state.

\begin{figure}
\includegraphics[scale=0.6,angle=0]{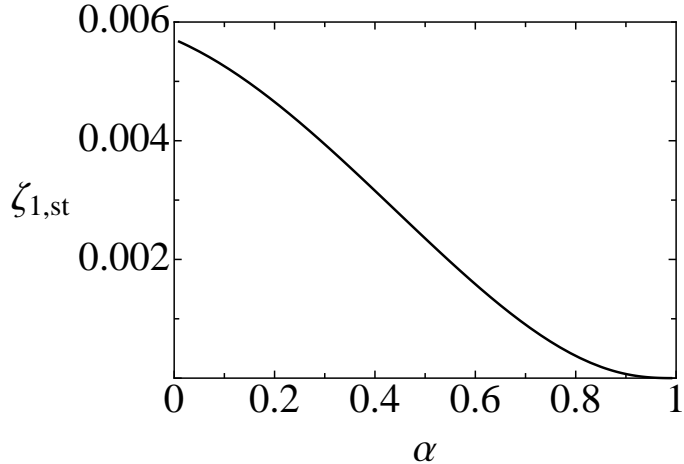}
\caption{Dimensionless Euler transport coefficient of the two-dimensional confined granular gas in the homogeneous steady state $\zeta_{1,st}$, as a function of the coefficient of normal restitution $\alpha$.}
\label{fig6}
\end{figure}

\begin{figure}
\includegraphics[scale=0.6,angle=0]{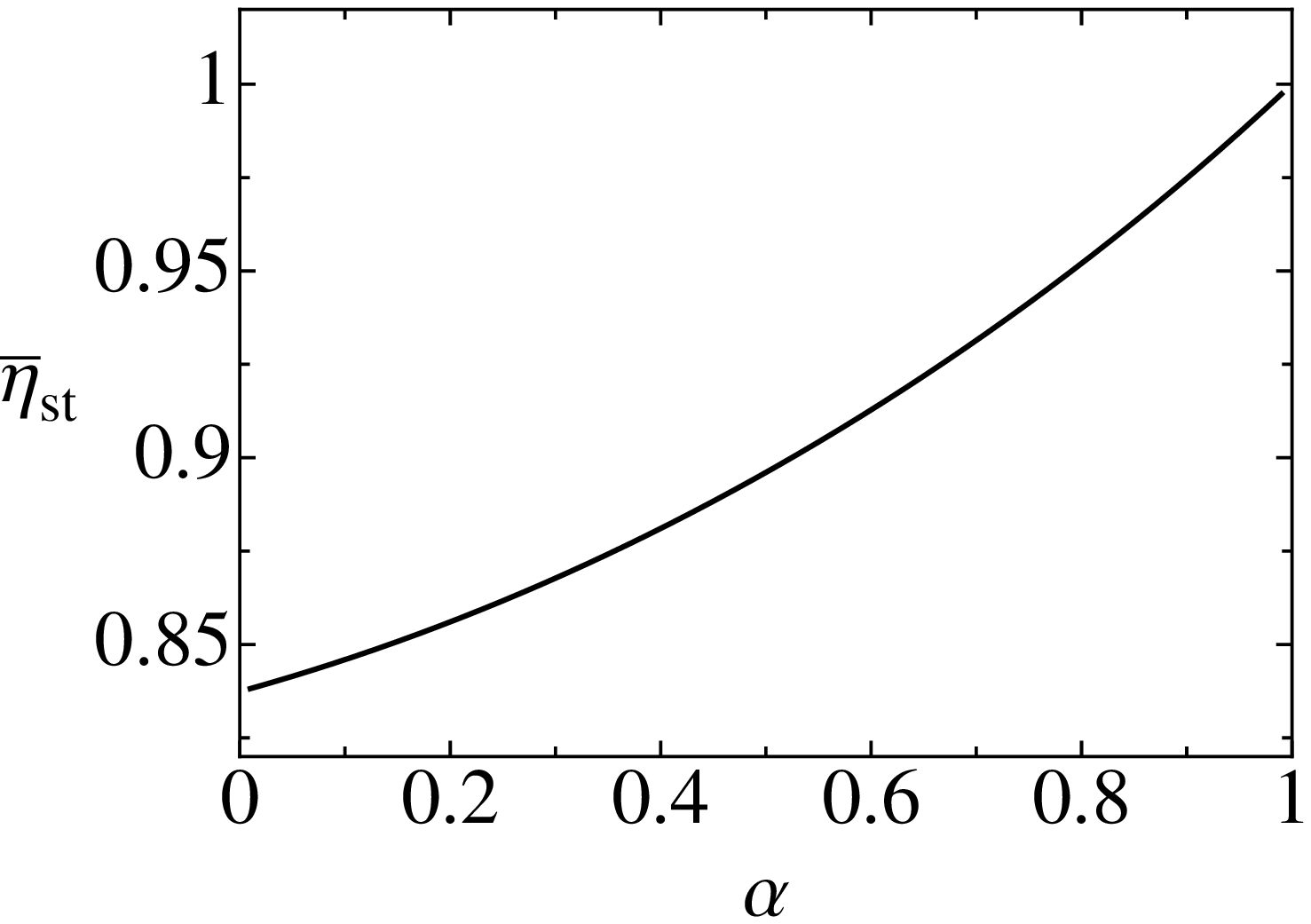}
\caption{Dimensionless shear viscosity of the two-dimensional confined granular gas in the homogeneous steady state $\overline{\eta}_{1,st}$, as a function of the coefficient of normal restitution $\alpha$.}
\label{fig7}
\end{figure}

\begin{figure}
\includegraphics[scale=0.6,angle=0]{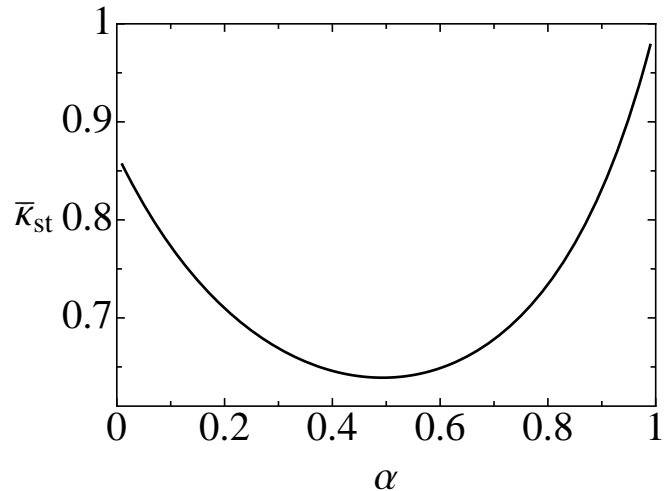}
\caption{Dimensionless (thermal) heat conductivity of the two-dimensional confined gas  in the homogeneous steady state $\overline{\kappa}_{st}$, as a function of the coefficient of normal restitution $\alpha$.}
\label{fig8}
\end{figure}

\begin{figure}
\includegraphics[scale=0.6,angle=0]{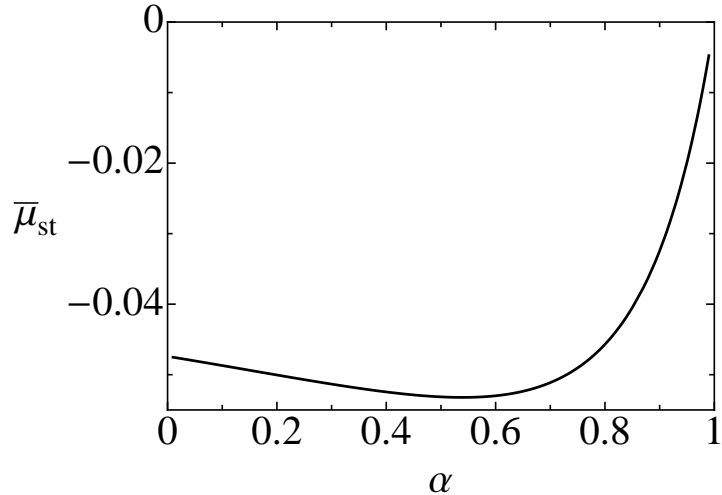}
\caption{ Dimensionless diffusive heat conductivity  of the two-dimensional confined granular gas in the homogeneous steady state $\overline{\mu}_{st}$, as a function of the coefficient of normal restitution $\alpha$.}
\label{fig9}
\end{figure}

\section{Discussion and conclusions}
\label{s5}
The objective of this work has been to derive the hydrodynamic equations to Navier-Stokes order for a model of confined granular gas from an underlying kinetic
theory, with all the parameters given explicitly. For clarification and context, the following comment must be taken into account. When applying the Chapman-Enskog procedure, the distribution function has been  computed up to first order in the gradients of the hydrodynamic fields density, flow velocity, and granular temperature. Consequently, the heat and momentum  fluxes are also determined to the same order. Since they occur as divergences in the balance equations, they lead to terms of second order in the gradients in those equations, what is usually referred to as the Navier-Stokes approximation for the fluxes. Also the rate of change of the temperature $\zeta$ has been computed to first order, but it appears without any gradient operator in front of it in the balance equation for the energy.  It follows that consistency of the Navier-Stokes order would require, in principle, computing  $\zeta$ up to second order in gradients, i.e. going an order further in the Chapman-Enskog expansion of the distribution function, the Burnett order. Such a calculation is rather involved and lengthy. The only case in which we are aware that second order contributions from $\zeta$ have been analyzed deals with the linear contributions in a low density gas of smooth inelastic hard spheres 
\cite{BDKyS98}. There, it is found than the terms are very small as compared with the similar ones arising from the fluxes. A similar behavior is likely to occur here with all the second order in the gradients contributions from $\zeta$.

The derived hydrodynamic equations are general, in the sense of having no restriction with regards to the values of the coefficient of normal restitution $\alpha$ or the velocity parameter of the model $\Delta$. In particular, they hold in principle arbitrarily far from the homogeneous steady state, as long as the system be near an homogeneous hydrodynamic state.  In the steady state, both parameters $\alpha$ and $\Delta$ determine the temperature of the system, that is not an arbitrary parameter anymore. Of course, the general hydrodynamic equations can be particularized for the steady state, as it has been actually done here, but it must be emphasized that the general form of the hydrodynamic equations can not be inferred from the equations derived for the steady situation.

The shear viscosity of the model in the steady state has been measured by using event driven molecular dynamics simulations. The transport coefficient was obtained from the decay rate of the transverse current \cite{BRyS13,SRyB15}. In a dense system ($n\sigma^{2}=0.4)$ it was found that a linear fit in $(1-\alpha)$ gives
\begin{equation}
\label{5.1}
\overline{\eta} \simeq 1.0512 \sqrt{\pi} \left[ 1-0.28(1-\alpha) \right].
\end{equation}
An expansion of Eq. (\ref{4.2}) in powers on $\alpha$ to first order gives
\begin{equation}
\label{5.2}
\overline{\eta} \simeq 1- \frac{17}{64} (1-\alpha).
\end{equation}
As expected, the low density theory developed here is not able to predict the prefactor in Eq.\ (\ref{5.1}). In dense systems, the collisional contributions to momentum transfer, and consequently to the shear viscosity play a fundamental role, and those effects are neglected at the level of the Boltzmann equation. However, if the lowest order inelasticity correction is considered, the results obtained in this paper are in good agreement with the simulation results for dense systems. A similar conclusion was reached in Ref.\ \cite{SRyB15}.

A relevant and recurrent question when deriving hydrodynamic equations from kinetic theory is to determine the context in which the equations apply. The small parameter in the Chapman-Enskog expansion is the ratio of the mean free path relative to the wavelength of the variation of the hydrodynamic fields. The mean free path is independent of  the time for the homogeneous hydrodynamics. Consequently, it seems sensible to conclude that the conditions for the Navier-Stokes order hydrodynamics of the model are the same as for usual granular gases of inelastic hard spheres or disks and also for elastic collisions, i.e. for sufficiently large space and time scales as compared with the mean free path and the inverse collision frequency.

\section{Acknowledgements}

This research was supported by the Ministerio de Econom\'{\i}a y Competitividad  (Spain) through Grant No. FIS2014-53808-P (partially financed by FEDER funds).

\appendix

\section{Chapman-Enskog solution}
\label{ap1}
To zeroth order in the gradients, the Chapman-Enskog expansion leads to
\begin{equation}
\label{ap1.1}
\partial_{t}^{(0)} f^{(0)} ({\bm v})= \int d{\bm v}_{1}\, \overline{T}_{0}({\bm v},{\bm v}_{1}) f^{(0)}({\bm v}) f^{(0)} ({\bm v}_{1}),
\end{equation}
while the balance equations to this order  become
\begin{equation}
\label{ap1.2}
\partial_{t}^{(0)}n=0, \quad \partial_{t}^{(0)}{\bm u}=0, \quad \partial_{t}^{(0)}T= - \zeta^{(0)} T.
\end{equation}
The lowest order rate of change of the temperture is given by
\begin{equation}
\label{ap1.3}
\zeta^{(0)} ({\bm r},t)= -\frac{2}{n({\bm r},t) T({\bm r},t) d}\, \omega[f^{(0)},f^{(0)}].
\end{equation}
It is worth to stress that the macroscopic fields are not expanded in the Chapman-Enskog method, so that the zeroth order distribution function already provides the exact actual macroscopic fields. Using Eqs.\ (\ref{ap1.2}), Eq. (\ref{ap1.1}) can be transformed into
\begin{equation}
\label{ap1.4}
 - \zeta^{(0)} T \frac{\partial f^{(0)}}{\partial T} =\int d{\bm v}_{1}\, \overline{T}_{0}({\bm v},{\bm v}_{1}) f^{(0)}({\bm v}) f^{(0)} ({\bm v}_{1}).
 \end{equation}
 Since $f({\bm r},{\bm v},t)$ and, therefore, $f^{(0)}({\bm r},{\bm v},t)$ are normal, the latter must have the scaled form
 \begin{equation}
 \label{ap1.5}
 f^{(0)} ({\bm r},{\bm v},t) =  n v_{0}^{-d}(T)f^{(0)*} \left( {\bm c}, \Delta^{*} \right),
 \end{equation}
 where
 \begin{equation}
 \label{ap1.6}
 v_{0}(T) \equiv \left[ \frac{2T({\bm r},t)}{m} \right]^{1/2}
 \end{equation}
 is a characteristic local thermal velocity. The dimensionless function $f^{(0)*}$ only depends on the temperature through the scaled velocity,
 \begin{equation}
 \label{ap1.8} 
 {\bm c} \equiv \frac{{\bm V}({\bm r},t)}{v_{0}({\bm r},t)}
 \end{equation}
 and the dimensionless speed parameter
 \begin{equation}
 \label{ap1.9}
 \Delta^{*} \equiv \frac{\Delta}{v_{0}({\bm r},t)}\, .
 \end{equation}
As a consequence, Eq.\ (\ref{ap1.4}) is equivalent to
 \begin{equation}
 \label{ap1.10}
 \frac{\zeta^{(0)}}{2} \left[ \frac{\partial}{\partial {\bm V}} \cdot \left( {\bm V} f^{(0)} \right) + \Delta^{*} \frac{\partial}{\partial \Delta^{*}}\, f^{(0)}  \right]=\int d{\bm v}_{1}\, \overline{T}_{0}({\bm v},{\bm v}_{1}) f^{(0)}({\bm v}) f^{(0)} ({\bm v}_{1}).
 \end{equation}
This equation formally coincides with the one describing the homogeneous hydrodynamics of the system in the low density limit \cite{BMGyB14}. It is important to realize that the zeroth order approximation in the Chapman-Enskog method is not a local version, both in space and time, of the distribution function of the homogeneous steady state eventually reached by the system \cite{BGMyB13,BRyS13}, but a local distribution generated from that describing the time-dependent homogeneous hydrodynamics. This is a relevant general issue when dealing with hydrodynamics around a non-equilibrium  state \cite{Lu06}. In  the second Sonine approximation, the zeroth order distribution function is approximated by
\begin{equation}
\label{ap1.11}
f^{(0)} ({\bm r},{\bm v},t) \simeq  \frac{n}{\left( \pi^{1/2} v_{0} \right)^{d}} e^{-c^{2}} \left[  1+ a_{2} (\Delta^{*}) S^{(2)}(c^{2}) \right],
\end{equation} 
where 
\begin{equation}
\label{ap1.12}
S^{(2)} (c^{2}) \equiv \frac{d(d+2)}{8} -\frac{d+2}{2}\, c^{2} +\frac{c^{4}}{2}.
\end{equation}
Neglecting nonlinear in $a_{2}$ terms, substitution of Eq.\ (\ref{ap1.11}) into Eq.\ (\ref{ap1.3}) yields
\begin{equation}
\label{ap1.13}
\zeta^{(0)}(T) \approx \frac{2^{3/2} \pi^{(d-1)/2}n \sigma^{d-1} v_{0}(T)}{\Gamma \left( d/2 \right) d} \left[ \frac{1 - \alpha^{2}}{2} \left( 1+ \frac{3 a_{2}}{16} \right) - \alpha \left( \frac{\pi}{2} \right) ^{1/2} \Delta^{*} - \left( 1- \frac{a_{2}}{16} \right) \Delta^{*2} \right].
\end{equation}
An equation for $a_{2}(\Delta^{*})$ is obtained by using this expression into the Boltzmann equation (\ref{ap1.10}), multiplication of the equation by $c^{4}$,  and later integration over the velocity ${\bm c}$. If the quadratic terms in $a_{2}$, as well as a term proportional to $a_{2} \partial a_{2} / \partial  \Delta^{*}$, are neglected, one gets \cite{BMGyB14}
\begin{equation}
\label{ap1.14}
\frac{\partial a_{2}}{\partial \Delta^{*}} =  \left[ \frac{4}{\Delta^{*}}+ \frac{4 A_{1}+ B_{1}}{A_{0} \Delta^{*}} \right] a_{2} + \frac{4}{\Delta^{*}} + \frac{ B_{0}}{A_{0} \Delta^{*}} ,
\end{equation}
with the several coefficients given by
\begin{equation}
\label{ap1.15}
A_{0} (\alpha, \Delta^{*})= (d+2) \left[ \frac{1- \alpha^{2}}{2} - \left( \frac{\pi}{2} \right)^{1/2}  \alpha \Delta^{*} - \Delta^{*2} \right],
\end{equation}
\begin{equation}
\label{ap1.16}
A_{1} (\alpha, \Delta^{*}) = \frac{(d+2)}{16} \left[ \frac{3(1- \alpha^{2})}{2}\, + \Delta^{*2} \right],
\end{equation}
\begin{eqnarray}
\label{ap1.17}
B_{0}(\alpha, \Delta^{*}) &=& (2 \pi )^{1/2} \left(1+2d+3 \alpha^{2} +4 \Delta^{*2}\right) \alpha \Delta^{*}-3+4 \Delta^{*4} + \alpha^{2}+ 2 \alpha^{4}  \nonumber  \\
&& -2d \left( 1-\alpha^{2}-2 \Delta^{*2} \right)+ 2 \Delta^{*2} \left( 1+6 \alpha^{2} \right),
\end{eqnarray}
\begin{eqnarray}
\label{ap1.18}
B_{1}(\alpha,\Delta^{*})& = & \left( \frac{\pi}{2} \right)^{1/2} \left[ 2 -2d(1-\alpha)+7 \alpha+3 \alpha^{3} \right] \Delta^{*}- \frac{1}{16} \left\{ 85 + 4 \Delta^{*4}-18 (3+2 \alpha^{2}) \Delta^{*2} \right.  \nonumber \\
&& \left. - \left( 32 +87 \alpha+ 30 \alpha^{3} \right) \alpha -2d \left[ 6 \Delta^{*2}-(1+\alpha) (31-15 \alpha) \right] \right\}.
\end{eqnarray}
The normal solution of Eq.\ (\ref{ap1.14}) has been analyzed in Ref. \cite{BMGyB14}, by solving numerically the differential equation for different initial conditions, and identifying the common part of all the generated solutions.  Independently of the approximation used to calculate it,  the solution of Eq. (\ref{ap1.10}) is isotropic in velocity space, i.e. it only depends on $V$. It follows that the lowest order pressure tensor and heat flux are
\begin{equation}
\label{ap1.19}
{\sf P}^{(0)} ({\bm r},t)= p({\bm r},t)  {\sf I}, \quad {\bm J}_{q}^{(0)} ({\bf r},t)=0,
\end{equation}
where ${\sf I}$ is the unit pressure tensor in $d$ dimensions, and $p({\bm r},t)$ is the hydrodynamic pressure
\begin{equation}
\label{ap1.20}
p({\bm r},t)= n({\bm r},t) T({\bm r},t).
\end{equation}
To first order in $\epsilon$, the expansion of the Boltzmann equation gives
\begin{equation}
\label{ap1.21}
\partial_{t}^{(0)} f^{(1)}+ {\cal L} f^{(1)} = - \partial_{t}^{(1)} f^{(0)}- {\bm v} \cdot \frac{\partial f^{(0)}}{\partial {\bm r}},
\end{equation}
with the linear operator ${\cal L}$ defined by
\begin{equation}
\label{ap1.22}
{\cal L} f^{(1)}({\bm r},{\bm v},t) \equiv - \int d {\bm v}_{1}\, \overline{T}_{0} ({\bm v},{\bm v}_{1}) \left[ f^{(0)}({\bm r},{\bm v},t) f^{(1)}({\bm r},{\bm v}_{1},t) +
f^{(1)}({\bm r},{\bm v},t) f^{(0)}({\bm r},{\bm v}_{1},t) \right].
\end{equation}
The macroscopic balance equations to this order are
\begin{equation}
\label{ap1.23}
\partial_{t}^{(1)} n+ {\bm u} \cdot {\bm \nabla} n+ n {\bm \nabla} \cdot {\bm u}=0,
\end{equation}
\begin{equation}
\label{ap1.24}
\partial_{t}^{(1)} {\bm u} + {\bm u} \cdot {\bm \nabla} {\bm u}+ (mn)^{-1} {\bm \nabla}p=0,
\end{equation}
\begin{equation}
\label{ap1.25}
\partial_{t}^{(1)} T+ {\bm u} \cdot {\bm \nabla} T + \frac{2T}{d} {\bm \nabla} \cdot {\bm u} = - \zeta^{(1)} T,
\end{equation}
with the first order in $\epsilon$ cooling rate being a linear functional of $f^{(1)}$,
\begin{equation}
\label{ap1.26}
\zeta^{(1)} \equiv -  \frac{4}{n({\bm r},t) T({\bm r},t) d} \omega \left[ f^{(0)},f^{(1)} \right].
\end{equation}
By using eqs. (\ref{ap1.23})-(\ref{ap1.25}), and that $f^{(0)}$ and $f^{(1)}$ are both normal distributions, Eq. (\ref{ap1.21}) is seen to be equivalent to
\begin{equation}
\label{ap1.27}
\partial_{t}^{(0)}f^{(1)}+{\cal L} f^{(1)} - \frac{\partial f^{(0)}}{\partial T}\, \zeta^{(1)} T = {\bm A} \cdot {\bm \nabla} \ln T +{\bm B} \cdot {\bm \nabla}  \ln n+ {\sf C} : {\bm \nabla} {\bm u},
\end{equation}
where
\begin{equation}
\label{ap1.28}
{\bm A} ({\bm V} | n,T)= - \frac{T}{m} \frac{\partial f^{(0)}}{\partial {\bm V}} + \frac{\bm V}{2} \frac{\partial }{\partial {\bm V}}\, \cdot \left( {\bm V} f^{(0)} \right) + \frac{\bm V}{2} \Delta^{*} \frac{\partial f^{(0)}}{\partial \Delta^{*}},
\end{equation}
\begin{equation}
\label{ap1.29}
{\bm B}({\bm V}| n,T) = - {\bm V} f^{(0)} - \frac{T}{m}\, \frac{\partial f^{(0)}}{\partial {\bm V}},
\end{equation}
\begin{equation}
\label{ap1.30}
{\sf C} ({\bm V}| n,T) = \frac{\partial}{\partial \bm V} \left( {\bm V} f^{(0)} \right) - \frac{1}{d} \left[ \frac{\partial}{ \partial \bm V}\, \cdot \left( {\bm V} f^{(0)} \right) + \Delta^{*} \frac{\partial f^{(0)}}{\partial \Delta^{*}} \right] {\sf I}.
\end{equation}
Because of the presence of the term involving $\Delta^{*}$ on the right hand side of Eq.\ (\ref{ap1.30}), the tensor ${\sf C}$ is not traceless, contrary to what happens in a system of elastic particles \cite{McL89} and also in a  system of inelastic hard spheres or disks \cite{BDKyS98}. The solution of the linear equation (\ref{ap1.27}) must have the form
\begin{equation}
\label{ap1.31}
f^{(1)} ({\bm V} |n,T)=  \bm{\mathcal  A} \cdot {\bm \nabla} \ln T + \bm{\mathcal B} \cdot {\bm \nabla} \ln T + {\mathcal C} : {\bm \nabla} {\bm u}.
\end{equation}
Consider the first order cooling rate $\zeta^{(1)}$ given by Eq. (\ref{ap1.26}). It is a scalar and, therefore, the only nonvanishing contribution to it has the form
\begin{equation}
\label{ap1.32}
\zeta^{(1)} = \zeta_{1} {\bm \nabla} \cdot {\bm u}
\end{equation}
with the Euler transport coefficient $\zeta_{1}$ given by
\begin{equation}
\label{ap1.33}
\zeta_{1}= - \frac{4}{n T d^{2}}\, \omega \left[ f^{(0)}, \Tr {\mathcal C} \right],
\end{equation}
where $ \Tr {\cal C}$ denotes the trace of the tensor ${\cal C}$.  When Eqs. (\ref{ap1.2}),  (\ref{ap1.31}), and (\ref{ap1.32}) are substituted into Eq. (\ref{ap1.27}), equations for $\bm{\mathcal A}$, $\bm{\mathcal B}$, and ${\mathcal C}$ are found by equating coefficients of the various gradients of the hydrodynamic fields,
\begin{equation}
\label{ap1.34}
- \zeta^{(0)} T \frac{\partial \bm{\mathcal A}}{\partial T}-T \frac{\partial \zeta^{(0)}}{\partial T}\, \bm{\mathcal A} + {\mathcal L} \bm{\mathcal A}= \bm{A},
\end{equation}
\begin{equation}
\label{ap1.35}
- \zeta^{(0)} \bm{\mathcal A}- \zeta^{(0)} T \frac{\partial \bm{\mathcal B}}{\partial T}+ {\mathcal L} \bm{\mathcal B} = \bm{B},
\end{equation}
\begin{equation}
\label{ap1.36}
-\zeta^{(0)} T \frac{\partial  {\mathcal C}}{\partial T} + {\mathcal L} {\mathcal C} -T \frac{\partial f^{(0)}}{\partial T} \zeta_{1} {\sf I}= {\sf C}.
\end{equation}
Next, let us analyze  the contribution to  the fluxes of first order in the gradients. The expression for  pressure tensor contribution can be expressed as
\begin{equation}
\label{ap1.37}
{\sf P}^{(1)}= m \int d{\bm V}\, {\bm V} {\bm V} f^{(1)}({\bm V} )= m \int d{\bm V}\, {\sf D} ({\bm V}) f^{(1)}({\bm V}),
\end{equation}
where
\begin{equation}
\label{ap1.38}
{\sf D} ({\bm V}) \equiv m \left( {\bm V} {\bm V} -\frac{1}{d}\ V^{2} {\sf I} \right),
\end{equation}
since $f^{(0)}$ gives, by construction, the correct exact value of the hydrodynamic fields and, consequently, the contribituions to them from $f^{(i)}$, $i=1,2, \cdots$, must vanish. Then, taking into account the isotropy of the tensors and that ${\sf D}$ is traceless, it follows that
\begin{equation}
\label{ap1.39}
{\sf P}^{(1)} = \int d{\bm V}\, {\sf D}({\bm V}) {\mathcal C} ({\bm V}): {\bm \nabla} {\bm u} = - \eta \left[ {\bm \nabla} {\bm u}+ ({\bm \nabla} {\bm u})^{+}- \frac{2}{d}\, {\bm \nabla} \cdot {\bm u} {\sf I} \right],
\end{equation}
with $({\bm \nabla} {\bm u})^{+}$ being the transposed of $ {\bm \nabla} {\bm u}$ and
\begin{equation}
\label{ap1.40}
\eta=- \frac{1}{d^{2}+d-2} \int d{\bm V}\, {\sf D}({\bm V}) :  {\mathcal C} ({\bm V}).
\end{equation}
Proceeding in a similar way, it is seen that the heat flux to first order in gradients reads
\begin{equation}
\label{ap1.41}
{\bm J}_{q}^{(1)}= - \kappa {\bm \nabla}T - \mu {\bm \nabla} n.
\end{equation}
The coefficients in this expression are given by
\begin{equation}
\label{ap1.42}
\kappa= - \frac{1}{Td} \int d{\bm V}\, {\bm S}({\bm V}) \cdot   \bm{\mathcal  A} ({\bm V}),
\end{equation}
{\begin{equation}
\label{ap1.43}
\mu= -\frac{1}{nd} \int d{\bm V}\, {\bm S}({\bm V}) \cdot  \bm{\mathcal B}  ({\bm V}),
\end{equation}
where
\begin{equation}
\label{ap1.44}
{\bm S} ({\bm V}) \equiv  \left( \frac{mV^{2}}{2}- \frac{d+2}{2}\, T \right) {\bm V}.
\end{equation}
Define the frequencies
\begin{equation}
\label{ap1.45}
\nu _\eta  \equiv \frac{\int d{\bf V}\,{\sf D}({\bf V}): {\cal L} {\mathcal C}({\bm V})}{\int
d{\bf V}\, {\sf D} ({\bf V}): {\mathcal C}({\bf V})},
\end{equation}
\begin{equation}
\label{ap1.46}
\nu _\kappa  \equiv \frac{\int d{\bf V}\,{\bf S}({\bf V})\cdot {\cal L} \bm{\mathcal A}({\bf V})}{
\int d{\bf V}\,{\bf S}({\bf V})\cdot \bm{\mathcal A}({\bf V})},
\end{equation}
\begin{equation}
\label{ap1.47}
 \nu_\mu  \equiv \frac{\int d{\bf V}\,{\bm S}({\bf V})\cdot {\cal L} \bm{\mathcal B}({\bf V})}{\int
d{\bf V}\,{\bf S}({\bf V})\cdot \bm{\mathcal B}({\bf V})}\, .
\end{equation}
By using these definitions and Eqs.\ (\ref{ap1.34})-(\ref{ap1.36}), it is easy to obtain first order differential equations obeyed by the transport coefficients,
\begin{equation}
\label{ap1.48}
\left( \zeta^{(0)} T \frac{\partial}{\partial T} - \nu_{\eta} \right) \eta = - \frac{1}{d^{2}+d-2} \int d{\bm V} {\sf D}({\bm V}) : {\sf C}({\bm V})= -nT,
\end{equation}
\begin{eqnarray}
\label{ap1.49}
\left[ \zeta^{(0)}T \frac{\partial}{\partial T}- \nu_{\kappa}+ \frac{\partial (T \zeta^{(0)})}{\partial T} \right] \kappa & = & \frac{1}{Td} \int d{\bm V} {\bm S}({\bm V}) \cdot {\bm A}({\bm V})  \nonumber \\
&=  &-\frac{(d+2)nT}{2m}\, \left( 1+2a_{2}+T \frac{\partial a_{2}}{\partial T} \right),
\end{eqnarray}
\begin{equation}
\label{ap1.50}
\left( \zeta^{(0)} T \frac{\partial}{\partial T} - \nu_{\mu} \right) \mu+\frac{\zeta^{(0)}T \kappa}{n} = \frac{1}{nd} \int d{\bm V} {\bm S}({\bm V}) \cdot {\bm B}({\bm V}) = -\frac{(d+2)T^{2}}{2m} \, a_{2}.
\end{equation}
Upon deriving the last equalities in each of the three above equations, the second Sonine approximation for $f^{(0)}$, Eq. (\ref{ap1.11}) has been employed.

\section{Evaluation of the frequencies}
\label{ap2}
Approximated expression for the functions $\nu_{\eta}$, $\nu_{\kappa}$, and $\nu_{\mu}$ have been obtained by using a Sonine expansion, truncated to lowest order. Since the collision operator commutes with the rotation operator, it follows from Eqs.\  (\ref{ap1.34})-(\ref{ap1.36}) that $ \bm{\mathcal A}$ and  $\bm{\mathcal B}$ must be isotropic functions of the velocity times ${\bm S}({\bm V})$. Moreover, only the traceless part of ${\mathcal C}$ is needed, and it must be the product of an isotropic function times  ${\sf D}({\bm V})$. Then, to lowest order in a Sonine expansion,
\begin{equation}
\label{ap2.1}
\bm{\mathcal A} ({\bm V}) \propto f_{M}(V) {\bm S}({\bm V}), \quad \bm{\mathcal B}({\bm V}) \propto f_{M}(V) {\bm S}({\bm V}), \quad \mathcal{C} ({\bm V}) - \frac{\sf I}{d} \Tr{\mathcal C} \propto f_{M}(V) {\sf D}({\bm V}),
\end{equation}
where $f_{M}$ is the Maxwellian distribution,
\begin{equation}
\label{ap2.2}
f_{M}(V) = n \left( \frac{m}{2 \pi T} \right)^{d/2} \exp \left(- \frac{mV^{2}}{2T} \right).
\end{equation}
With this approximation,
\begin{equation}
\label{ap2.3}
\nu _\eta =\frac{\int d{\bf V}\,{\sf D}({\bf V}) : {\cal L} [f_M(V)
{\sf D}({\bf V})]}{\int d{\bf V}\,f_M(V){\sf D}({\bf V}) : {\sf D}({\bf V})}
=\frac{\int d{\bf V}\,D_{ij}({\bf V})
L[f_M({\bf V})D_{ij}({\bf V})]}{(d-1)(d+2)n T^{2}},
\end{equation}
\begin{equation}
\label{ap2.4}
\nu _\kappa =\nu _\mu =\frac{\int d{\bf V}\,{\bf S}({\bf V})\cdot
{\cal L}[f_M(V){\bf S}({\bf V})]}{\int d{\bf V}\,f_M({\bf V})
{\bf S}({\bf V})\cdot {\bf S}({\bf V})}
=\frac{2m \int d{\bf V}\,{\bf S}({\bf V})\cdot {\cal L}[f_M({\bf V})
{\bf S}({\bf V})]}{d(d+2)nT^{3}}.
\end{equation}
The evaluation of the integrals in the above expressions is straightforward, and it is facilitated by using symbolic computer programs. Since similar calculations have been reported many times in the literature, we merely report here the final results,
\begin{eqnarray}
\label{ap2.5}
\nu_{\eta}  & = & \frac{\sqrt{2} \pi^{\frac{d-1}{2}} n \sigma^{d-1} v_{0}(T)}{d (d+2) \Gamma \left(d/2\right)}\, \left[ \left( 1+\alpha \right) (2d+3-3\alpha) \left( 1- \frac{a_{2}}{32} \right)  \right. \nonumber \\\
&& \left. + \sqrt{2 \pi}  (d-2 \alpha) \Delta^{*}-2 \left( 1+ \frac{3a_{2}}{32} \right)  \Delta^{*2} \right],
\end{eqnarray}
\begin{eqnarray}
\label{ap2.6}
\nu_{\kappa}= \nu_{\mu} & = & \frac{\pi^{\frac{d-1}{2}}    n \sigma^{d-1}v_{0}(T)}{\sqrt{2} d(d+2) \Gamma \left(d/2 \right)} \left\{ 
\frac{(1+\alpha) \left[ 512+352d-96 \alpha (d+8) \right]}{2^{6}}  \right. \nonumber \\ 
&&+  \left.  \frac{(1+\alpha) [ 5d+4-3(4-d) \alpha]}{2^{6}}\, 
 a_{2} \right. \nonumber \\
&& - \left( \frac{\pi}{2} \right)^{1/2} \left[ 2 (1-d) + (d+8) \alpha \right] \Delta^{*}   \nonumber \\
&&- \left. \frac{32(d+8)-3(4-d)a_{2} }{2^{5}}\, \Delta^{*2}  \right\} .
\end{eqnarray}
For $\Delta^{*}=0$, the above expressions reduce to those obtained in Refs. \cite{BDKyS98} and \cite{ByC01}.

\section{The Euler transport coefficient}
\label{ap3}
The transport coefficient $\zeta_{1}$ is given by Eq. (\ref{ap1.33}). An equation for it can be obtained from Eq.\ (\ref{ap1.36}),
\begin{equation}
\label{ap3.1}
-\zeta^{(0)} T \frac{\partial  \Tr {\mathcal C} }{\partial T} +  {\mathcal L} \Tr {\mathcal C}- T \frac{\partial f^{(0)}}{\partial T} \zeta_{1} d= \Tr{\sf C}= - \Delta^{*}  \frac{\partial f^{(0)}}{\partial a_{2}} \frac{\partial a_{2}}{\partial \Delta^{*}}.
\end{equation}
Since $\Tr {\mathcal C}$ must have vanishing velocity moments up to second degree, its lowest order Sonine approximations reads
\begin{equation}
\label{ap3.2}
 \Tr {\mathcal C} \simeq    \frac{b_{2} e^{-c^{2}} }{\pi^{d/2} \sigma^{d-1}v_{0} ^{d+1}}\,   S^{(2)}(c^{2}).
 \end{equation}
To determine the dimensionless coefficient $b_{2}$, we substitute the above expression into Eq.\ (\ref{ap3.1}), and afterwards multiply it by   $v^{4}$ and integrate over ${\bm v}$. After some lengthy but trivial algebra, a differential equation is obtained,
\begin{equation}
\label{ap3.3}
\overline{\zeta}^{(0)}  \Delta^{*} \frac{\partial b_{2}}{\partial \Delta^{*}} - \left[ 3 \overline{\zeta}^{(0)} + \frac{8 \chi (\Delta^{*})}{d(d+2)} + 4 d \zeta_{1} (a_{2}+1) -d \zeta_{1} \Delta^{*} \frac{\partial a_{2}}{\partial \Delta^{*}} \right] b_{2}=-2 \Delta^{*} \frac{\partial a_{2}}{\partial \Delta^{*}}.
\end{equation}
Here,
\begin{eqnarray}
\label{ap3.4}
\overline{\zeta}_{1}\equiv \frac{\zeta_{1}}{b_{2} } & = & - \frac{4 \pi^{- \frac{d+1}{2}}}{d^{2}} \int d{\bm c}_{1} \int d{\bm c}_{2}\, e^{-c_{1}^{2}-c_{2}^{2}} \left[1+a_{2} S^{(2)}(c_{1}^{2}) \right] S^{(2)}(c_{2}^{2}) \nonumber \\
&& \times  \left[ \frac{\Delta^{*2} c_{12}}{\Gamma \left( \frac{d+1}{2} \right) } + \frac{\pi^{1/2} \alpha \Delta c_{12}^{2} }{2\Gamma \left( \frac{d+2}{2} \right)} -\frac{(1-\alpha^{2}) c_{12}^{3}}{4 \Gamma \left (\frac{d+3}{2}  \right)} \right]
\end{eqnarray}
and
\begin{eqnarray}
\label{ap3.5}
\chi (\Delta^{*})  & \equiv & - \frac{1}{b_{2} v_{0}^{4} n}\, \int d{\bm v} v^{4} {\mathcal L} \Tr{\mathcal  C} \nonumber \\
&= & \frac{1}{b_{2} v_{0}^{4} n }\,
\int d{\bm v} \int d{\bm v}_{1}\, f^{(0)}({\bm r},{\bm v},t) \Tr{\mathcal  C}({\bm v}_{1}) T_{0} ({\bm v}, {\bm v}_{1}) (v^{4}+v_{1}^{4}).
\end{eqnarray}
The integrals in the above expressions of $\zeta_{1} $ and $\chi$ can be carried out getting
\begin{equation}
\label{ap3.6}
\overline{\zeta}_{1}= \frac{\pi^{\frac{d-1}{2}}}{2^{\frac{17}{2}} d^{2} \Gamma \left( d/2 \right)} \left[ 96+9a_{2}-3 \alpha^{2} \left( 32+3a_{2} \right)+ \Delta^{*2} (64+30a_{2}) \right],
\end{equation}
\begin{eqnarray}
\label{ap3.7}
\chi  &= & \frac{\pi^{\frac{d-1}{2}}}{2^{11}  \Gamma \left( d/2 \right)} \left( \sqrt{2} \left\{ 30 \alpha^{4} (32-a_{2})-5 (544+7a_{2})-4 \Delta^{*2} (32+15 a_{2}) \right. \right. 
\nonumber \\
& &- 64(d-1) \alpha (16+a_{2})  -2d (992+17a_{2} )+ 3 \alpha^{2} \left[ 928+43 a_{2} \right. \nonumber \\
& & \left. \left. +12 \Delta^{*2}(32+3a_{2}) +10 d (32-a_{2}) \right] + 6 \Delta^{*2} (288-45a_{2}+64d+6d a_{2} ) \right\}  \nonumber \\
& & \left. +512 \sqrt{\pi} \Delta^{*} \left[ 2+7 \alpha +3 \alpha^{3} -2d(1-\alpha) \right] \right).
\end{eqnarray}
Now, the dimensionless coefficient $b_{2}$ is obtained by solving numerically the set of equations (\ref{ap1.14}) and (\ref{ap3.3}), and identifying the hydrodynamic part of the solution as discussed in the main text for the shear viscosity. Afterwards, the Euler transport coefficient $\zeta_{1}$ follows by using Eq.\ (\ref{ap3.4}).


\begin{thebibliography}{99}
\bibitem{d} Although for the particular system considered here it is $d=2$, we formulate the theory for both $d=2$ and $d=3$, because of the possible applications of similar models to other contexts.
\end{thebibliography}
\end{document}